\documentclass[aps,pra,reprint,showpacs,superscriptaddress, nofootinbib]{revtex4-2}

\usepackage{graphicx}% Include figure files
\usepackage{dcolumn}% Align table columns on decimal point
\usepackage{bm}
\usepackage{xcolor}  
\usepackage[normalem]{ulem}
\expandafter\let\csname equation*\endcsname\relax
\expandafter\let\csname endequation*\endcsname\relax
\usepackage[dvipsnames,usenames]{xcolor}
\usepackage{amsmath}
\usepackage{graphicx}
\usepackage{soul}
\usepackage{float}
\usepackage{physics}
\usepackage{amsfonts}
\emergencystretch=\maxdimen
\usepackage[export]{adjustbox}
\usepackage[plainpages=false,pdfpagelabels,colorlinks=true,linkcolor=red,urlcolor=PineGreen,citecolor=PineGreen,pdftitle={Title},pdfauthor={},pdfdisplaydoctitle=true,pdfduplex=DuplexFlipLongEdge]{hyperref}
\usepackage{comment}

\providecommand{\Tr}{\operatorname{Tr}}
\providecommand{\Re}{\operatorname{Re}}
\usepackage{orcidlink}

\begin{document}

\title{Stochastic Quantum Information Geometry and Speed Limits at the Trajectory Level}

\author{Pedro \surname{B.~Melo}\,\orcidlink{0000-0002-3726-761X}}
\email{pedrobmelo@aluno.puc-rio.br}
 \affiliation{Universit\`a degli Studi di Palermo, Dipartimento di Fisica e Chimica - Emilio Segr\`e, via Archirafi 36, I-90123 Palermo, Italy}
 \affiliation{Departamento de F\'isica, PUC-Rio, 22452-970, Rio de Janeiro RJ, Brazil}
 \author{Pedro \surname{V.~Paraguass\'u}\,\orcidlink{0000-0003-2334-5688}}
 \affiliation{Departamento de F\'isica, PUC-Rio, 22452-970, Rio de Janeiro RJ, Brazil}
 \author{S\'ilvio \surname{M.~Duarte~Queir\'os}\,\orcidlink{0000-0001-6915-1892}}
 \affiliation{Centro Brasileiro de Pesquisas F\'isicas, 22452-970, Rio de Janeiro RJ, Brazil}
 \author{Fernando \surname{Iemini}\,\orcidlink{0000-0003-1645-9492}}
 \affiliation{Departamento de F\'isica, Universidade Federal Fluminense, 22452-970, Niter\'oi RJ, Brazil} 
  \author{Mauro \surname{Paternostro}\,\orcidlink{0000-0001-8870-9134}}
 \affiliation{Universit\`a degli Studi di Palermo, Dipartimento di Fisica e Chimica - Emilio Segr\`e, via Archirafi 36, I-90123 Palermo, Italy}
 \affiliation{Centre for Quantum Materials and Technologies, School of Mathematics and Physics, Queen’s University Belfast, BT7 1NN, United Kingdom}
 \author{Welles \surname{A.~M.~Morgado}\,\orcidlink{0000-0002-9690-6258}}
 \affiliation{Departamento de F\'isica, PUC-Rio, 22452-970, Rio de Janeiro RJ, Brazil}
\date{\today}

\begin{abstract}
In quantum metrology, precision is typically characterized by an ensemble-averaged quantity, the quantum Fisher information (QFI), which averages over the fluctuations of individual measurement records. Here we introduce the conditional quantum Fisher information (CQFI), a trajectory-level version of the QFI that generalizes the classical stochastic Fisher information to the quantum domain. Defined through the symmetric logarithmic derivative and conditioned on a measurement outcome, the CQFI is a random variable whose average recovers the QFI. Using it, we derive a trajectory-level quantum speed limit, illustrated by the quantum-jump unraveling of a driven thermal qubit. Moreover, the CQFI decomposes into incoherent (population) and coherent (basis-rotation) contributions, together with an interference cross-term. This cross-term vanishes on average but can take negative values along single trajectories, providing a local witness of destructive interference between classical and quantum information channels.
\end{abstract}
 
\maketitle

\section{Introduction}

The Quantum Fisher Information (QFI) stands as a cornerstone of quantum metrology, quantifying the ultimate sensitivity of a quantum system to changes in an unknown parameter. By establishing the quantum Cramér-Rao bound, the QFI dictates the fundamental precision limits for parameter estimation, constrained only by the laws of quantum mechanics~\cite{Paris2009, Tóth_2014, Liu_2020, MONTENEGRO20251, escher2011general}. This framework has proven indispensable across diverse physical systems, from single-qubit sensors~\cite{Ramsey1950, Wineland1998} to complex many-body quantum simulators~\cite{Giovannetti2011, Pezze2018}.

Traditionally, the QFI is employed to assess precision based on ensemble-averaged probability distributions obtained from repeated measurements. This ensemble approach underpins key protocols such as quantum state tomography~\cite{G_Mauro_D_Ariano_2002} and optical phase estimation in noisy interferometry~\cite{Genoni2011, Demkowicz2012}. More recently, the QFI has found profound utility in the study of quantum critical phenomena, where its divergence near critical points signals that criticality can be harnessed to dramatically enhance metrological precision, a resource known as critical metrology~\cite{Zanardi_2006, Zanardi_2008, Frerot2018, Louis2020, Chu2021}. Furthermore, the QFI has provided a rigorous geometric basis for deriving quantum speed limits (QSLs) in open systems, bounding the rate of evolution for non-unitary processes subject to environmental decoherence~\cite{Taddei2013, delCampo2013, Deffner2013}.

Parallel developments in classical stochastic thermodynamics have established the Fisher Information (FI) as a fundamental metric for thermodynamic state spaces. Beyond its metrological origins, the FI has emerged as a geometric measure connecting information theory to thermodynamics, particularly for near-equilibrium states~\cite{Crooks2007, Crooks2009, Sivak2012}. The thermodynamic length and action, derived from the FI metric, have been shown to relate directly to excess work and dissipation in driven systems. These results have been generalized to arbitrarily far-from-equilibrium systems~\cite{Ito2018, Nicholson2018}, enabling the derivation of fundamental speed limits for the evolution of both states and observables in classical stochastic processes~\cite{Ito2020, Nicholson2020, GarciaPintos2022}.

Connecting information geometry with stochastic thermodynamics, the stochastic Fisher information (SFI) has been introduced to quantify the {\it surprisal} rate of individual trajectories allowing speed limits for state transformations at the level of single realizations~\cite{melo2_2025}.This trajectory-level perspective has proven particularly valuable for understanding rare events and non-typical fluctuations that are masked in ensemble averages~\cite{Touchette2009, Garrahan2018}. However, a fully quantum generalization that accounts for the intricate interplay between classical population statistics and quantum coherence evolution along individual measurement trajectories has remained an open challenge. Here we address this problem by introducing the conditional quantum Fisher information (CQFI).

This extension, however, is far from straightforward, as the quantum case introduces difficulties absent in the classical setting. Quantum measurements inevitably disturb the system state, leading to trajectory-dependent evolution that differs qualitatively from classical stochastic processes~\cite{Wiseman2009, Jacobs2014}. {Moreover, the} quantum superposition principle introduces correlations between population dynamics and coherent evolution that have no classical analog. {Beyond this, the} choice of measurement basis fundamentally affects the observed trajectory statistics, requiring a framework that can account for arbitrary measurement schemes while preserving the connection to fundamental quantum limits.

The CQFI is constructed to meet precisely these requirements. Defined through the symmetric logarithmic derivative (SLD) and conditioned on a specific measurement outcome, it generalizes the classical SFI to the quantum domain while preserving its trajectory-level interpretation, thereby extending quantum metrology from ensemble averages to single-shot realizations. Unlike the standard QFI, which provides a global bound averaged over all possible measurement outcomes, the CQFI is a random variable that fluctuates along individual quantum trajectories.

\begin{figure}[H]
    \centering
    \includegraphics[width=8.6cm]{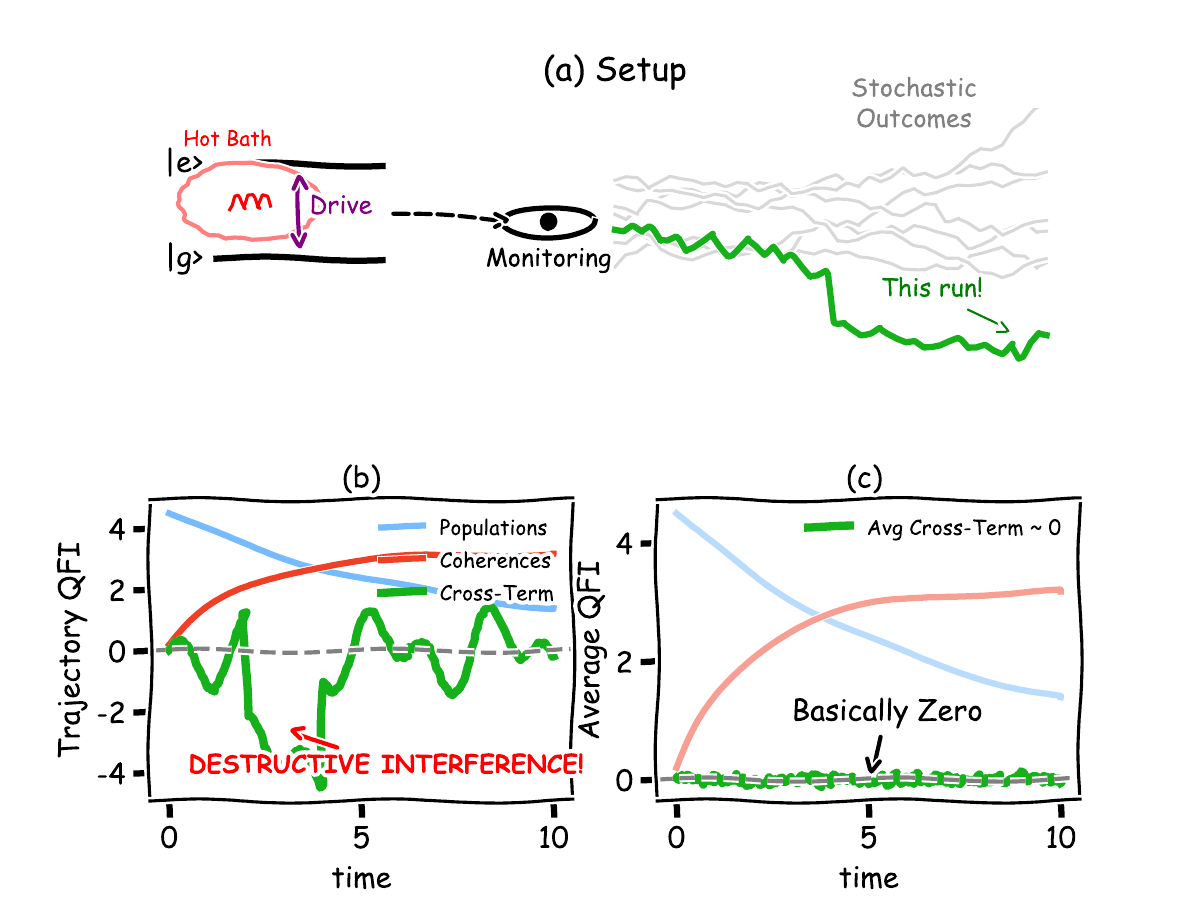}
    \caption{Sketch of the conditional quantum Fisher information for the adopted example in this work. Panel (a) shows a sketch of a qubit in contact with a thermal bath, submitted to a driving potential. When the system is monitored, it\ produces stochastic outcomes for each realization. Panel (b) depicts the results for trajectory level quantum Fisher information decomposed into the contributions from population changes in blue, the contributions from unitary rotations of the eigenbasis in red, and the cross contributions from the interference between both channels in green. Panel (c) shows the results from the averaged QFI on ensemble level, where the cross contributions vanish.}
    \label{fig:Cartoon_CQFI}
\end{figure}

Our approach yields several contributions of relevance for quantum metrology and stochastic thermodynamics. First, we demonstrate that the CQFI admits a physically transparent decomposition into three distinct contributions: (i) an \emph{incoherent} term arising from population changes, analogous to the classical SFI; (ii) a \emph{coherent} term arising from unitary rotations of the eigenbasis; and (iii) a \emph{cross-term} representing interference between classical and quantum channels. We show that the latter  term can be negative, signifying destructive interference between the evolutions of population and coherences along individual trajectories, which is a purely quantum phenomenon that disappears at ensemble level. Figure \ref{fig:Cartoon_CQFI} depicts this interference effect in a sketch, showing that on trajectory level negative contributions to the CQFI can appear, while on ensemble average they are zero. 

Second, leveraging this decomposition, we construct a stochastic information geometry that defines thermodynamic length and action for individual quantum trajectories. This geometric framework allows us to quantify the statistical distance traversed by a quantum system in a single experimental realization, extending the tools of information geometry from ensemble descriptions to the realm of individual quantum measurements.

Third, we derive fundamental quantum speed limits valid at the single-trajectory level, generalizing both classical stochastic bounds~\cite{melo2_2025} and ensemble quantum speed limits~\cite{Taddei2013, delCampo2013} to the intersection of these domains. We demonstrate that such trajectory-level bounds can be significantly tighter than their ensemble counterparts, particularly in regimes dominated by unlikely-to-occur but informative quantum trajectories.

Because the Conditional Quantum Fisher Information preserves the global operator structure while assigning an information value to each measurement outcome, it is an appropriate quantity to cope with the quantum challenges aforementioned since it manages to live at the single-shot level while remaining anchored to ensemble quantum limits. In other words, the CQFI  separates the information carried by population changes from that carried by basis rotations and, crucially, exposes a transient interference term that captures the nonclassical coupling between these channels along individual trajectories. On the other hand, since the CQFI is SLD-based, it is defined from the ensemble density matrix and thus links single‑trajectory diagnostics to ensemble geometry, enabling a stochastic information metric, trajectory action and length, and trajectory‑level quantum speed limits that remain consistent with -- and in general tighter than -- their ensemble counterparts \cite{Ruppeiner1995}.

Finally, we validate our theoretical framework through detailed numerical simulations using two paradigmatic systems: the quantum jump unraveling of a driven thermal qubit and the continuous monitoring of displaced Gaussian states. These examples demonstrate that our trajectory-level bounds provide useful insights into quantum information dynamics that are not accessible through ensemble-averaged approaches.

The remainder of this paper is organized as follows. Sec.~\ref{secII} introduces the CQFI formalism and derives its spectral decomposition into incoherent, coherent, and interference contributions. Sec.~\ref{secIII}  applies the CQFI framework to construct a stochastic information geometry with time as the estimation parameter. Sec.~\ref{sec:SLs} derives trajectory-level quantum speed limits and presents comprehensive numerical validations using quantum jump trajectories. In Sec.~\ref{secV} we draw our conclusions, which include a discussion of applications in adaptive quantum metrology, real-time quantum feedback, and single-trajectory thermodynamic protocols. A set of appendices report the technical aspects of our derivations. 

\section{Conditional Quantum Fisher Information}
\label{secII}

Consider a physical system described by a probability distribution $p(x|\theta)$, where $x$ represents an accessible physical parameter of the system (such as position, momentum, or a measurement outcome) and $\theta$ is an unknown parameter to be estimated. The classical Fisher information (FI) for $\theta$ is defined as the expectation value of the squared score function (the derivative of the log-likelihood)
\begin{equation}
    \mathcal{F}(\theta) = \sum_{x} p(x|\theta) \left( \frac{\partial \log p(x|\theta)}{\partial \theta} \right)^2.
\end{equation}

An analogous quantity is the Stochastic Fisher information (SFI), recently introduced in the context of stochastic thermodynamics \cite{Melo3_2025,melo2_2025}. The SFI is a random variable whose average yields the FI. Unlike the standard FI, the SFI $\iota(x,\theta)$ depends on the specific single-shot realization $x$
\begin{equation}
    \iota(x,\theta) = \left( \frac{\partial \log p(x|\theta)}{\partial \theta} \right)^2.\label{eq:SFI}
\end{equation}
The SFI effectively quantifies the ``surprisal'' rate with respect to the parameter $\theta$. Concretely, the surprisal of a trajectory is $-\log p(x_{[0,t]}|\theta)$; its sensitivity to $\theta$ is the score $\partial_\theta \log p(x_{[0,t]}|\theta)$, whose square is the SFI \eqref{eq:SFI}. The stochastic thermodynamic interpretation of the SFI closely mirrors that of the FI, maintaining the analogy between fluctuating (trajectory-level) and averaged quantities. Furthermore, the SFI provides speed limits for state transformations at the single-trajectory level \cite{melo2_2025}.

\subsection{Quantum Fisher information}

The quantum Fisher information (QFI) generalizes the FI to quantum systems \cite{Paris2009}. We define the conditional probability distribution via the Born rule as $p(x|\theta) = \mathrm{Tr}(\Pi_x \rho_\theta)$, where $\Pi_x = \ketbra{x}{x}$ is a projector onto the state $\ket{x}$, and $\rho_\theta$ is a quantum state in a Hilbert space $\mathcal{H}$ parameterized by $\theta \in \mathcal{M}$ (a $d$-dimensional manifold). This definition generalizes to cases where $\{\Pi_x\}$ forms an arbitrary positive operator-valued measure (POVM).

In contrast to the classical setting, the Chentsov-Petz theorem establishes that the choice of Fisher metric for density matrices is not unique \cite{Toth2013}. Using the spectral decomposition $\rho_\theta = \sum_{x} p_x(\theta) \ketbra{x(\theta)}{x(\theta)}$, the general QFI is given by
\begin{equation}
    \mathcal{F}_{Q}(\theta) = \sum_{x,y} \frac{|\bra{x(\theta)}\partial_{\theta}\rho_\theta\ket{y(\theta)}|^2}{p_{x}(\theta) f(p_{y}(\theta)/p_{x}(\theta))},
\end{equation}
where $f$ is an operator monotone function (self-inversive $f(t) = t f(1/t)$ and normalized $f(1)=1$). We focus on the Symmetric Logarithmic Derivative (SLD) QFI, corresponding to $f_{\mathrm{SLD}}(t) = (t + 1)/2$. In this case, $\mathcal{F}_{Q}(\theta)$ simplifies to
\begin{equation}
    \mathcal{F}_{Q}(\theta) = 2 \sum_{x,y} \frac{|\bra{x(\theta)}\partial_{\theta}\rho_\theta\ket{y(\theta)}|^2}{p_{x}(\theta) + p_{y}(\theta)},
\end{equation}
where the sum runs over indices with $p_x + p_y > 0$. Unless otherwise stated, we refer to the SLD-based quantity simply as the QFI.

We introduce the SLD operator $L_{\theta}$, implicitly defined by the Lyapunov equation
\begin{equation}
    \frac{\partial\rho_\theta} {\partial\theta} = \frac{1}{2} \{ L_{\theta}, \rho_\theta \},
\end{equation}
where $\{ \cdot, \cdot \}$ denotes the anti-commutator. In the spectral basis of $\rho_\theta$, $L_{\theta}$ takes the form
\begin{equation}
\begin{aligned}
    L_{\theta} &= \sum_{x} \frac{\partial_{\theta}p_{x}(\theta)}{p_{x}(\theta)}\ketbra{x(\theta)}{x(\theta)} \\
    & + 2\sum_{x \neq y} \left(\frac{p_x(\theta) - p_{y}(\theta)}{p_{x}(\theta) + p_{y}(\theta)}\right) \braket{x(\theta)}{\partial_{\theta}y(\theta)} \ketbra{x(\theta)}{y(\theta)}. \label{eq:SLD_form}
\end{aligned}
\end{equation}
The QFI is then given by the second moment of the SLD operator
    $\mathcal{F}_{Q}(\theta) = \mathrm{Tr}(\rho_\theta L_{\theta}^2)$.
Replacing Eq.~(\ref{eq:SLD_form}) in the trace, we obtain the decomposition $\mathcal{F}_{Q}(\theta)= \mathcal{F}_Q^{\mathrm{IC}}(\theta)+ \mathcal{F}_Q^{\mathrm{C}}(\theta)$ with
\begin{equation}
\begin{aligned}
    \mathcal{F}_Q^{\mathrm{IC}}(\theta)&= \sum_{x} \frac{(\partial_{\theta}p_x(\theta))^2}{p_x(\theta)},\\
     \mathcal{F}_Q^{\mathrm{C}}(\theta)&=2 \sum_{x \neq y} \sigma_{xy} |\braket{y(\theta)}{\partial_{\theta}x(\theta)}|^2,
    \label{eq:QFI_Eigenvectors}
\end{aligned}
\end{equation} 
where we have introduced the coefficients
   $ \sigma_{xy} = \frac{(p_x(\theta) - p_y(\theta))^2}{p_x(\theta) + p_y(\theta)}$~\cite{Paris2009}.
This decomposition identifies an incoherent contribution to the QFI, associated with $\mathcal{F}_Q^{\mathrm{IC}}(\theta)$ and stemming from a change in the populations, and a coherent one encompassed by $\mathcal{F}_{Q}^{\mathrm{C}}(\theta)$, which arises  from the rotation of the eigenvectors with respect to the instantaneous basis.

\subsection{Conditional Quantum Fisher Information}

We now introduce the quantum generalization of the SFI, which we dub \textit{Conditional Quantum Fisher Information} (CQFI). Recall that the SFI describes the information content of a single measurement outcome purely from the trajectory statistics. In the quantum regime, we must account for how the global state structure dictates local sensitivity.

Let $\{\Pi_\alpha\}$ be a general POVM, where $\alpha$ labels the measurement outcome. The conditional probability is $p(\alpha|\theta) = \mathrm{Tr}(\Pi_{\alpha}\rho_\theta)$. From Eq.~\eqref{eq:SFI}, the SFI associated with outcome $\alpha$ is 
\begin{equation}
    \iota(\alpha,\theta) = \left( \frac{\partial \log p(\alpha|\theta)}{\partial \theta} \right)^2 = \left( \frac{\mathrm{Tr}(\Pi_{\alpha}\partial_\theta \rho_\theta)}{\mathrm{Tr}(\Pi_{\alpha}\rho_\theta)} \right)^2.
\end{equation}
The expression above is an exact expression for the surprisal rate, but it depends on the specific outcome only through scalar traces and does not take on the local SLD structure. Bounding it via $\Re(z)^2 \le |z|^2$ and the Cauchy–Schwarz inequality [cf. Appendix \ref{app:cqfi}], we define the CQFI for a specific outcome $\alpha$ as the resulting upper bound,
\begin{equation}
     f_{Q,\alpha}(\theta) = \frac{\mathrm{Tr}(\Pi_\alpha L_{\theta} \rho_\theta L_\theta)}{\mathrm{Tr}(\Pi_\alpha \rho_\theta)},
\end{equation}
\textit{i.e.} $\iota(\alpha,\theta) \le f_{Q,\alpha}(\theta)$. Physically, the CQFI represents the QFI conditioned on a specific POVM element. It quantifies the local power of the SLD generator sampled by the outcome $\alpha$. Note that while $f_{Q,\alpha}(\theta)$ is a random variable associated with a single realization, it relies on the operator $L_\theta$, which is defined globally via the ensemble density matrix $\rho_\theta$.

The CQFI naturally averages to the total QFI over the measurement statistics. Using the completeness relation $\sum_\alpha \Pi_\alpha = \mathbb{I}$, we have
\begin{equation}
\begin{aligned}
    \langle f_{Q,\alpha}(\theta)\rangle_{\{\Pi_\alpha\}} &= \sum_\alpha p(\alpha|\theta) f_{Q,\alpha}(\theta)   \\&= \sum_\alpha \mathrm{Tr}(\Pi_\alpha L_\theta \rho_\theta L_\theta)  \\& = \mathrm{Tr}\left[ \left(\sum_\alpha\Pi_\alpha\right) L_{\theta} \rho_\theta L_\theta \right] \\&  = \mathrm{Tr}( L_{\theta} \rho_\theta L_\theta) = \mathcal{F}_Q(\theta),
\end{aligned}
\end{equation}
where the third equality uses the completeness relation $\sum_\alpha \Pi_\alpha = \mathbb{I}$ and the last uses the cyclicity of the trace, $\mathrm{Tr}(L_\theta\rho_\theta L_\theta) = \mathrm{Tr}(\rho_\theta L_\theta^2)$.
Another interesting definition is the state-conditioned SLD variance, defined by $\tilde{f}_{Q,\alpha}(\theta) = \text{Tr}(\Pi_\alpha L_\theta^2)$. While this is mostly not equivalent to the CQFI [cf. Appendix \ref{app:cqfi}.~4] , a key advantage of this quantity is that it allows for the clear identification of contributions stemming from the interplay between classical statistics and quantum coherence. Assuming $\Pi_\alpha=\ketbra{\alpha}$, the projected SLD variance $\tilde{f}_{Q,\alpha}$ can be written as (cf. Appendix~\ref{App:B})
\begin{equation}
\label{trajdec}
     \tilde{f}_{Q,\alpha}(\theta) = \tilde{f}_{Q,\alpha}^{IC}(\theta) + \tilde{f}_{Q,\alpha}^{C}(\theta) + \tilde{f}_{Q,\alpha}^{X}(\theta).
\end{equation}
The first two terms are the incoherent and coherent contribution, respectively, analogous to the ensemble-based expression 
\begin{align}
     \tilde{f}_{Q,\alpha}^{IC}(\theta) & = \sum_n\left(\frac{\partial_\theta p_n}{p_n}\right)^2\abs{\braket{n}{\alpha}}^2,\label{eq:CQFI_IC}\\
     \tilde{f}_{Q,\alpha}^{C}(\theta) & = \sum_k \left|\sum_{n(\ne k)}\braket{n}{\alpha}\frac{2(p_n - p_k)}{p_n + p_k}\braket{k}{\partial_\theta n}\right|^2.\label{eq:CQFI_C}
\end{align}
Crucially, at the single-outcome level, the cross-term $\tilde{f}_{Q,\alpha}^X$ appears in Eq.~\eqref{trajdec}. Explicitly, we have that 
\begin{align}
     \tilde{f}_{Q,\alpha}^{X}(\theta)& = \sum_{k\ne n} \text{Re}\left[\braket{\alpha}{k(\theta)}\braket{n(\theta)}{\alpha}\left(\frac{\partial_\theta p_k}{p_k} + \frac{\partial_\theta p_n}{p_n}\right) \right. \nonumber \\
    & \left.\times \left(\frac{2(p_n - p_k)}{p_n + p_k}\right) \braket{k}{\partial_\theta n}\right],\label{eq:CQFI_X}
\end{align}
thus  showing that such genuinely trajectory-dependent contribution arises from the interference between population shifts and basis rotations. Unlike the incoherent and coherent terms, which are strictly non-negative, the cross-term $\tilde{f}_Q^X$ can be negative.  While, at first sight, this might appear counterintuitive, negative values imply destructive interference, where the classical re-weighting of probabilities and the quantum rotation of the basis act in opposition relative to the probe state $\ket{\alpha}$. This interference effect is a strictly local feature: when averaged over the ensemble, the cross-term vanishes ($\langle \tilde{f}_Q^X \rangle = 0$), allowing us to recover the standard decomposition of the QFI. If the measurement basis coincides with the spectral basis of $\rho_\theta$ (i.e., $\ket{\alpha} = \ket{x}$), the off-diagonal terms drop out and $\tilde{f}_{Q,x}(\theta) = \tilde{f}_{Q,x}^{\mathrm{IC}}(\theta)$; if, in addition, $L_\theta|x\rangle$ has support only within the eigenspace of $p_x$ [cf. Appendix~\ref{app:equality}, condition~(i)], this coincides with the primary CQFI $f_{Q,x}(\theta)$ and recovers the classical SFI limit. This is the case, for example, under quasi-static or adiabatic evolutions, where the system remains in an eigenstate.

This embodies a key central result of our study: the CQFI generalizes the SFI to the quantum domain, preserving the decomposition into probability evolution and basis rotation, while unveiling the transient interference hidden in ensemble averages. To approach the quantum thermodynamic geometry \cite{roadmap2025} conditioned to the trajectory level, in the following Section, we make use of the CQFI with respect to time as a metric.

\section{Conditional Quantum Fisher Information with respect to Time}
\label{secIII}

The estimation of time is not limited to clock synchronization protocols; it also enables a rigorous connection between the CQFI, speed limits, and stochastic thermodynamics. To this end, we hereby construct the CQFI formalism with time as the estimated parameter, setting the stage for the discussion on speed limits that follows.

\subsection{Classical framework}
Let us consider a path defined by a set of discrete probability distributions $\{p_{x}(t)\}$ over discrete states $x \in X$, evolving over a time interval from $t = 0$ to $t = \tau$. Assuming the system is controlled by a finite set of time-dependent parameters $\theta(t) = (\theta_1(t),\dots,\theta_{M}(t))$, the path is confined to a statistical manifold $\Theta = \{p(x|\theta(t))\}$. Treating time $t$ as the parameter of interest (\textit{i.e.} $\theta_1(t) = \theta_2(t) = \dots = \theta_M(t) = t$), the Fisher information (FI) is given by
\begin{equation}
    \mathcal{I}(t) = \sum_{x} p_{x}(t) \left(\frac{\mathrm{d}\log p_{x}(t)}{\mathrm{d}t}\right)^2.
\end{equation}
The FI induces a metric structure on the manifold, given by the line element
\begin{equation}
    \mathrm{d}s^2 = \frac{1}{4}\mathcal{I}(t)\mathrm{d}t^2.
\end{equation}
This allows for the definition of the thermodynamic length
\begin{equation}
    \mathcal{L}(t) = \frac{1}{2}\int_{0}^{t} \mathrm{d}\tau \sqrt{\mathcal{I}(\tau)},
\end{equation}
which represents a distance on the statistical manifold. The derivative $\mathrm{d}s/\mathrm{d}t = \frac{1}{2}\sqrt{\mathcal{I}(t)}$ expresses the instantaneous statistical speed. The statistical divergence, or thermodynamic action, is defined as
\begin{equation}
    \mathcal{J}(t) = \frac{t}{4}\int_{0}^{t}\mathrm{d}\tau~\mathcal{I}(\tau).
\end{equation}
This quantity is analogous to a kinetic energy integral and serves as a bound to the squared statistical distance via the Cauchy-Schwarz inequality, $\mathcal{J}(t) \ge \mathcal{L}^2(t)$.

When considering single realizations of the dynamics, the stochastic Fisher information (SFI) with respect to time adopts the form~\cite{melo2_2025}
\begin{equation}
    \iota(x,t) = \left(\frac{\mathrm{d}\log p_{x}(t)}{\mathrm{d}t}\right)^2.
\end{equation}
Given that the underlying dynamics are stochastic, $\iota(x,t)$ satisfies the requirements for a random metric on the statistical manifold. The stochastic length $\ell[x(t)]$ for a specific path (sequence of states $x(t)$) is defined as
\begin{equation}
    \ell[x(t)] = \frac{1}{2} \int_{0}^{t} \mathrm{d}\tau \sqrt{\iota(x,\tau)}.
\end{equation}
Similarly, the stochastic divergence $j[x(t)]$ is given by
\begin{equation}
    j[x(t)] = \frac{t}{4}\int_{0}^{t} \mathrm{d}\tau ~\iota(x,\tau),
\end{equation}
which bounds the squared stochastic velocity according to $j[x(t)] \ge \ell^2[x(t)]$.

\subsection{Quantum framework}

This framework generalizes to the quantum setting by considering the CQFI of Eqs.(\ref{eq:CQFI_IC})--(\ref{eq:CQFI_X}) with time $t$ as the estimation parameter. Previous studies have analyzed this decomposition at the ensemble level, where the cross-terms vanish upon averaging, leaving the total QFI as the sum of an incoherent contribution $\mathcal{F}_{Q}^{\mathrm{IC}} = \lim_{N_{{\rm trajs}}\rightarrow \infty}\langle \tilde{f}_{Q}^{\mathrm{IC}}\rangle$ and a coherent contribution $\mathcal{F}_{Q}^{\mathrm{C}} = \lim_{N_{{\rm trajs}}\rightarrow \infty}\langle \tilde{f}_{Q}^{\mathrm{C}}\rangle$. The term $\mathcal{F}_{Q}^{\mathrm{C}}$ serves as a measure of coherence utility. Recently, Bettmann and Goold~\cite{Bettmann2025} utilized this ensemble decomposition to provide a thermodynamic interpretation of $\mathcal{F}_{Q}^{\mathrm{IC}}$ and derived bounds on the entropic velocity using $\mathcal{F}_{Q}^{\mathrm{C}}$.

Our work establishes a trajectory-level approach to these quantities. To do so, we employ the quantum trajectory formalism, specifically the quantum jump method (or Monte Carlo Wave Function method) \cite{Manzano2022}.

We assume the system evolves under a Gorini-Kossakowski-Sudarshan-Lindblad (GKSL) master equation
\begin{equation}
    \frac{\mathrm{d}\rho_{t}}{\mathrm{d}t} = -\frac{i}{\hbar}[H_{t}, \rho_{t}] + \mathcal{D}[\rho_{t}],
\end{equation}
with the dissipator 
\begin{equation}
    \mathcal{D}[\rho_{t}] = \sum_{k} \left(L_{k}\rho_{t} L_{k}^{\dagger} - \frac{1}{2}\{L_{k}^{\dagger}L_{k}, \rho_{t}\}\right).
\end{equation}
Here, $\{L_{k}\}$ is a set of jump operators satisfying the detailed balance condition $L_{k_-} = L_{k_+}e^{-\Delta s_{k}/2}$, with $L_{k_+} = \sqrt{\Gamma_+}L^\dagger_k$, $L_{k_-} = \sqrt{\Gamma_-}L_k$ and $\Delta s_k$ represents the stochastic entropy flow from the system to the environment, associated to the reversible part of entropy changes due to the interaction between the system and the environment \cite{Horowitz_2013, Elouard2017}. While the GKSL equation describes the ensemble dynamics, we unravel this master equation into individual quantum trajectories. A single trajectory is described by a pure state $\ket{\psi_{\gamma}(t)}$, labeled by the realization index $\gamma$. The ensemble state is recovered via the statistical average $\rho_{t} = \mathbb{E}[\ketbra{\psi_{\gamma}(t)}{\psi_\gamma(t)}]$.

The stochastic Schrödinger equation (SSE) governing the evolution of a single trajectory is
\begin{equation}
\begin{aligned}
    \mathrm{d}\ket{\psi_{\gamma}(t)} &= \left[-\frac{i}{\hbar}H_{\mathrm{eff}}(t) + \frac{1}{2}\sum_{k}\| L_{k}\ket{ \psi_{\gamma}(t)}\|^2\right]\ket{\psi_{\gamma}(t)}\mathrm{d}t \\
    & + \sum_{k}\left[\frac{L_{k}}{\|L_{k}\ket{\psi_{\gamma}(t)}\|} - 1\right]\ket{\psi_\gamma(t)}\mathrm{d}N_{k}(t),
\end{aligned}
\end{equation}
where $H_{\mathrm{eff}}(t) = H(t) - \frac{i\hbar}{2}\sum_{k}L_{k}^{\dagger}L_{k}$ is the non-Hermitian effective Hamiltonian. The term $\mathrm{d}N_{k}(t)$ represents a Poisson process increment such that $\mathrm{d}N_k=1$ if a jump occurs and $0$ otherwise, with ensemble average $\mathbb{E}[\mathrm{d}N_k] = \|L_k \ket{\psi_\gamma}\|^2 \mathrm{d}t$.

The probability of observing a specific trajectory $\gamma_{[0,\tau]} = \{n_0, \gamma_{(0,\tau)}, n_\tau\}$ is given by $P_{\Lambda}(\gamma_{[0,\tau]}) = p_{n_0}^0 \mathrm{Tr}[\Pi_{n_\tau}^{\tau} \mathcal{T}_{\Lambda}(\gamma_{[0,\tau]})\Pi_{n_0}^{0}\mathcal{T}_{\Lambda}^{\dagger}(\gamma_{[0,\tau]})]$, where the trajectory is bookended by projective measurements $\Pi_{n}^0$ and $\Pi_{n}^\tau$ in the eigenbasis of the system. Here the closed interval $\gamma_{[0,\tau]}$ denotes the full record including the bookend measurement outcomes $n_0$ and $n_\tau$, whereas the open interval $\gamma_{(0,\tau)}$ denotes the deterministic no-jump segment generated by $H_{\mathrm{eff}}$ between consecutive jumps. The dynamics thus has a renewal structure: each jump resets the conditioned state, and the no-jump segments $\gamma_{(0,\tau)}$ are the inter-jump waiting periods of the underlying point process.

We define the stochastic length for a single trajectory $\gamma$ as
\begin{equation}
    \ell(\gamma,t) = \frac{1}{2}\int_{0}^{t} \mathrm{d}\tau\sqrt{f_{Q,\gamma}(\tau)},
\end{equation}
and the single-trajectory action as
\begin{equation}
    j(\gamma,t) = \frac{t}{4}\int_{0}^{t}\mathrm{d}\tau~f_{Q,\gamma}(\tau).
\end{equation}
Unlike the ensemble statistical action, which relates to a geodesic distance on the statistical manifold, the single-trajectory action $j(\gamma,t)$ does not strictly represent a divergence from a geodesic path, as geodesics are ill-defined for the stochastic metric itself. However, $j(\gamma,t)$ remains a useful quantity due to the Cauchy-Schwarz inequality $j(\gamma,t) \ge \ell^2(\gamma,t)$. Furthermore, as shown in classical stochastic thermodynamics \cite{melo2_2025}, the hierarchy of speed limits holds on average $\mathcal{J}(t) \ge \mathcal{L}^2(t) \ge \mathrm{Var}\{\ell[\gamma]\}$.

We emphasize that the three-way decomposition of Eqs. (\ref{eq:CQFI_IC})–(\ref{eq:CQFI_X}) is a property of the projected SLD variance $\tilde{f}_{Q,\alpha}$, whereas the stochastic length and action below are built from the normalized CQFI $f_{Q,\gamma}$. For a pure trajectory state the latter reduces to the Fubini–Study speed squared [cf. Eq. (32) and surrounding discussion], and the two objects coincide under the equality condition of Appendix A.4; the incoherent/coherent/cross structure should therefore be read as diagnosing the content of the trajectory's sensitivity, while $f_{Q,\gamma}$ is the quantity the speed limits bound.

To characterize the stochastic information geometry at the trajectory level, we observe that, at each time step $t$, the CQFI is conditioned on the instantaneous pure state of the trajectory $\ket{\psi_\gamma(t)}$. By making the spectral decomposition of the ensemble density matrix we can determine the projection of the state at a given time for one trajectory onto each of the eigenstates of the system $\braket{n(t)}{\psi_\gamma(t)}$.

\section{Quantum Speed Limits on the Trajectory Level \label{sec:SLs}}

Having introduced the formalism of quantum trajectories, we can now define stochastic dynamics to develop the quantum analogue of speed limits for single trajectories~\cite{melo2_2025}. Significant progress has been made in understanding quantum speed limits (QSLs) for open system dynamics and their thermodynamic interpretation. For instance, QSLs have been derived for open systems~\cite{delCampo2013, Taddei2013} and generalized to infinite families of metrics~\cite{Pires2016}. Also, thermodynamic interpretations have been given to the Fisher information~\cite{Nicholson2018,Ito2018,Nicholson2020} and, more recently, to the QFI~\cite{Bettmann2025}. Here, we extend the stochastic results of Ref.~\cite{melo2_2025} to the quantum regime and the ensemble-level results of Ref.~\cite{Bettmann2025} to the trajectory level.

From the Cauchy-Schwarz inequality applied to the stochastic geometric quantities, $j[\rho_{\gamma}(t)] \ge \ell^{2}[\rho_{\gamma}(t)]$, it follows that the time-averaged variance of the CQFI is non-negative~\cite{Nicholson2018,Bettmann2025}
\begin{equation}
    \delta(\gamma_{[0,\tau]}) = 4\frac{j[\rho_{\gamma}(t)] - \ell^2[\rho_{\gamma}(t)]}{t^2} \ge 0,
\end{equation}
where $j[\rho_{\gamma}(t)]$ and $\ell[\rho_{\gamma}(t)]$ are the accumulated action and length up to time $t$. Due to their stochastic nature along a single trajectory, it is impossible to define a geodesic in the same sense as the ensemble QFI geometric quantities $\mathcal{J}$ and $\mathcal{L}$. However, we can reconstruct non-geodesic inequalities at the trajectory level. Given the positivity of $j[\rho_{\gamma}(t)]$, $\ell^2[\rho_{\gamma}(t)]$, and $\delta[\rho_{\gamma}(t)]$, we have the following bound
\begin{equation}
    \delta[\rho_{\gamma}(t)] \le \frac{4j[\rho_{\gamma}(t)]}{t^2}.
\end{equation}
Consequently, we can establish the trajectory-level  inequality 
\begin{equation}
    \frac{\mathcal{I}[\rho_{\gamma}(t)]}{\delta[\rho_{\gamma}(t)]} \ge 1,
\end{equation}
where $\mathcal{I}[\rho_{\gamma}(t)] = \frac{1}{t}\int_{\gamma} \mathrm{d}\tau f_{Q}[\rho_{\gamma}(\tau)]$ is the time-averaged CQFI along the trajectory $\gamma$ (defined for $0 \le \tau \le t$).

Previous works~\cite{GarciaPintos2022, Bringewatt2024, Bettmann2025} have shown that the SLD operator connects the time-derivative of the density matrix to the QFI via the bound
\begin{equation}
    \int_0^t \mathrm{d}t' \frac{|\dot{o}(t')|}{\Delta_{\rho_{t'}}O} \le \int_0^t \mathrm{d}t' \sqrt{\mathcal{F}_Q(t')} = 2\mathcal{L}(t).\label{eq:Avg_QSLs_Obs_QFI}
\end{equation}
Here, $|\dot{o}(t)| = |\mathrm{Tr}[O \dot{\rho}_t]|$ represents the rate of change of the observable expectation value, and $\Delta_{\rho_t}O = \sqrt{\mathrm{Tr}[\rho_t O^2] - \mathrm{Tr}[\rho_t O]^2}$ is the instantaneous variance of the observable $O$.

We propose analogous bounds for the instantaneous state of a single trajectory, $\rho_\gamma(t) = \ketbra{\psi_\gamma(t)}{\psi_\gamma(t)}$. We define the stochastic rate of change of the observable as $\dot{o}_\gamma(t) = \mathrm{Tr}[O \dot{\rho}_\gamma(t)]$ and the stochastic variance with respect to the instantaneous pure state as $\Delta_{\rho_\gamma(t)}O$. The new single-trajectory speed limit is derived as
\begin{equation}
    |\dot{o}_\gamma(t)| \le \Delta_{\rho_\gamma(t)} O \sqrt{f_Q[\rho_\gamma(t)]}.
\end{equation}
The integral version of this bound yields
\begin{equation}
    \int_0^t \mathrm{d}t' \frac{|\dot{o}_\gamma(t')|}{\Delta_{\rho_\gamma(t')} O} \le \int_0^t \mathrm{d}t' \sqrt{f_Q[\rho_\gamma(t')]} = 2\ell[\rho_\gamma(t)].\label{eq:Trajs_QSLs_Obs_QFI}
\end{equation}
This inequality represents a fundamental speed limit at the trajectory level. To validate these bounds, we apply them to the jump unraveling of a driven two-level system in contact with a thermal reservoir.

\subsection*{Example: Driven two-level system in a thermal environment}

We test these results using a single two-level system coupled to a thermal environment at finite temperature $T$. Here $T$ denotes the absolute temperature of the bath, which fixes the mean thermal occupation $\bar{n} = (e^{\hbar\omega/k_B T} - 1)^{-1}$ entering the jump rates below ($k_B$ being the Boltzmann constant). The system Hamiltonian is $H_{S} = \omega\sigma_{z}$, subject to a time-dependent driving potential $V(t) = \varepsilon (e^{-i \omega t}\sigma_{+} + e^{i\omega t}\sigma_{-})$, where $\varepsilon \ll \omega$ is the driving amplitude and $\{\sigma_{z}, \sigma_+, \sigma_-\}$ are the Pauli matrices. The control parameter is defined by the phase factor $\lambda(t) = e^{i\omega t}$. 
\begin{figure}[H]
    \centering
    \includegraphics[width=8.6cm]{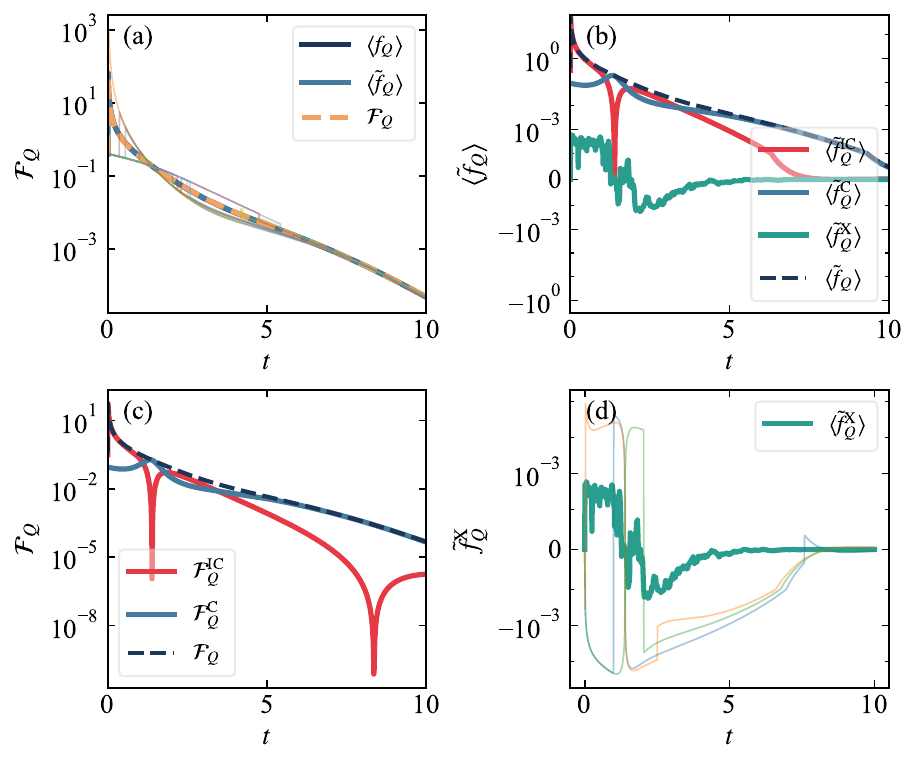}
    \caption{Decomposition and validation of the CQFI. (a) The agreement of the average CQFI $\langle f_Q\rangle$ (in blue) with the QFI $\mathcal{F}_Q$ (in orange dashed), with relative error $\le 1\%$, along with the trajectory CQFI $f_Q$. (b) Decomposition of the projected SLD variance into populations $\langle \tilde{f}_Q^{{\rm IC}}\rangle$, coherent $\langle \tilde{f}_Q^{{\rm C}}\rangle$, and (trajectory-level) cross $\tilde{f}_Q^{{\rm X}}$ contributions. The coherence term dominates at longer times, highlighting the quantum nature of the dynamics. (c) Decomposition of $\mathcal{F}_Q$ (black dashed) into populations $\mathcal{F}_Q^{{\rm IC}}$ (in red) and coherences $\mathcal{F}_Q^{{\rm C}}$ (in blue) contributions, where the cross term is strictly zero. (d) Trajectory level fluctuations of $\tilde{f}_Q^X$ (in green, for $N_{{\rm traj}} = 3$) against its average behavior.}
    \label{fig:CQFI_driven}
\end{figure}
We assume a cyclic protocol $\Lambda$ with duration $\tau$, such that $\lambda(0) = \lambda(\tau)$. In the rotating wave approximation, the Hamiltonian becomes $\hat{H}_{RWA} = \varepsilon\hat{\sigma}_x$. As such, the Hamiltonian is time-independent in this frame. 

The dynamics are described by the GKSL master equation with jump operators corresponding to the emission and absorption of photons
\begin{equation}
    L_- = \sqrt{\Gamma_0 (\bar{n} + 1)} \sigma_-; \quad L_{+} = \sqrt{\Gamma_0 \bar{n}}\sigma_+,
\end{equation}
where $\Gamma_0$ is the spontaneous emission rate and $\bar{n} = (e^{\hbar\omega/k_B T} - 1)^{-1}$ is the mean number of thermal photons. For a continuously monitored system, the stochastic Schrödinger equation is given by
\begin{align}
    \mathrm{d}\ket{\psi_{\gamma}(t)} &= \mathrm{d}t \left[-i H_{\mathrm{eff}}(t) + \frac{\Gamma_0}{2}(\langle \sigma_+ \sigma_-\rangle - \sigma_+ \sigma_-)\right]\ket{\psi_{\gamma}(t)} \nonumber\\
    &+ \mathrm{d}N_{-}\left(\frac{\sigma_-}{\sqrt{\langle\sigma_+\sigma_-\rangle}} - 1\right) \ket{\psi_\gamma(t)}\nonumber\\
    &+ \mathrm{d}N_{+}\left(\frac{\sigma_+}{\sqrt{\langle\sigma_-\sigma_+\rangle}} - 1\right) \ket{\psi_\gamma(t)},
\end{align}
where $H_{\mathrm{eff}}$ includes the driving and non-Hermitian damping terms. The Poissonian increments $\mathrm{d}N_{\pm}$ take the value $1$ when a jump $k = \pm$ occurs and $0$ otherwise.
\begin{figure}[H]
    \centering
    \includegraphics[width=8.6cm]{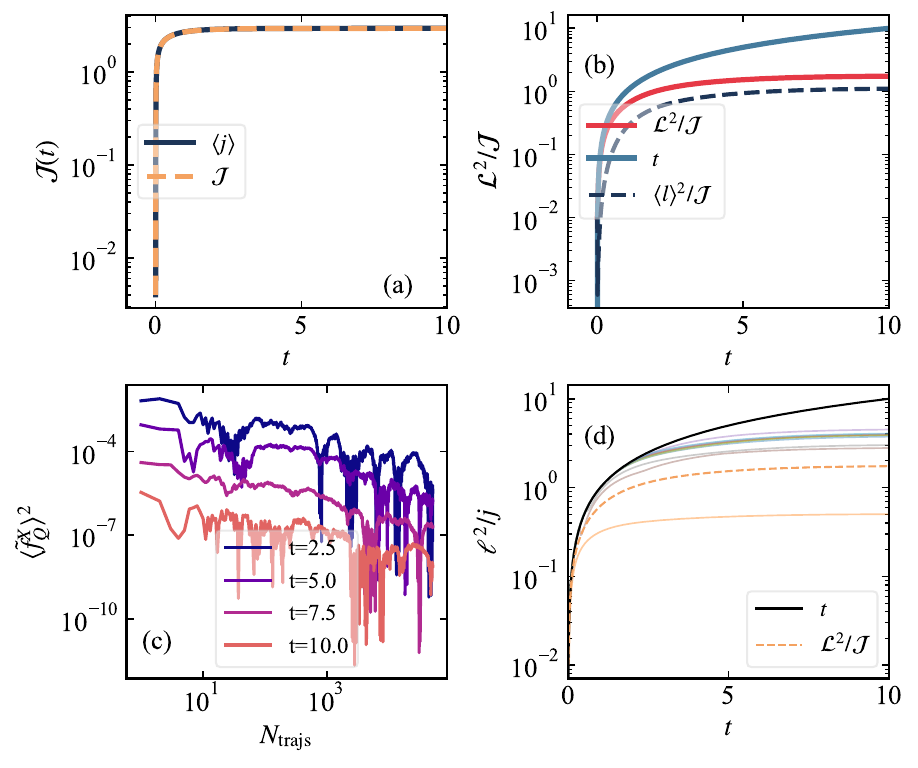}
    \caption{Verification of information geometry and speed limits. (a) Convergence of the average stochastic action $\langle j\rangle(t)$ to the ensemble-level statistical action $\mathcal{J}(t)$. (b) Comparison of speed limits at ensemble and averaged-trajectory level. The bound derived for ensemble QFI is consistently higher than or equal to average of single-trajectories bounds, respecting the expected hierarchy. (c) Convergence of $\langle \tilde{f}_Q^{{\rm X}}\rangle$ toward zero as the number of trajectories increases, for different time steps. (d) Verification of the trajectory-level speed limit inequality for a subset of the $N_{{\rm trajs}} = 5\times10^4$ simulated trajectories. 
    }
    \label{fig:QSLs_driven}
\end{figure}
We proceed by calculating the stochastic length $\ell_{\Lambda}(\gamma_{{[0,\tau]}})$ for each trajectory. Given the probability overlap with the ensemble $c_n(\gamma(t)) =   \braket{n(t)}{\psi_\gamma(t)}$, the projected SLD variance for this two-level system takes the specific form
\begin{align}
     \tilde{f}_{Q}^{\mathrm{IC}}(\gamma(t)) & = \sum_{n = 0,1} |c_n(\gamma(t))|^2 \left(\frac{\dot{p}_{n}(t)}{p_{n}(t)}\right)^2, \\
     \tilde{f}_{Q}^{\mathrm{C}}(\gamma(t)) & = \sum_{k= 0,1} \left| c_n(\gamma(t)) \frac{2(p_n - p_k)}{p_n + p_k} \braket{k(t)}{\dot{n}(t)} \right|^2_{n\ne k},\\
     \tilde{f}_{Q}^{\mathrm{X}}(\gamma(t)) & = \sum_{k\ne n} \text{Re}\left[c_k^*(\gamma(t))c_n(\gamma(t))\left(\frac{\dot{p}_k(t)}{p_k} + \frac{\dot{p}_n(t)}{p_n}\right) \right. \nonumber \\
    & \quad \times \left. \left(\frac{2(p_n - p_k)}{p_n + p_k}\right) \braket{k}{\dot{n}(t)}\right].
\end{align}
The incoherent and coherent terms $\tilde{f}_Q^{{\rm IC}}$, and $\tilde{f}_Q^{{\rm C}}$ are diagonal with relation to the projections on state $\ketbra{\psi_\gamma(t)}$, and the cross term $\tilde{f}_{Q}^{{\rm X}}$ contains the off-diagonal projections. 

The numerical simulations validate our framework, demonstrating that the average CQFI converges to the standard QFI with negligible error, as seen in Fig. \ref{fig:CQFI_driven}(a). However, the power of the trajectory-level approach becomes apparent in the spectral decomposition [Fig. \ref{fig:CQFI_driven}(b)-(c)]. Unlike the ensemble QFI, the single-trajectory decomposition reveals a sign-changing cross-term $\tilde{f}_Q^X$ arising from the interference between population changes and basis rotations. At the parameters used here ($\varepsilon = 0.15$, $\Gamma_0 = 0.4$, $T = 0.8$), the cross-term is comparable to the coherent term and sign-changing, rather than strictly order-one. As illustrated in Fig. \ref{fig:CQFI_driven}(d), this term often exhibits negative fluctuations during transient evolution, signaling a destructive interplay between classical and quantum information channels that is strictly non-classical and unobservable in standard ensemble averages.

Fig.~\ref{fig:QSLs_driven} shows the numerical consistency of stochastic information geometry. We find that the trajectory-averaged action numerically recovers the standard ensemble statistical action, while the speed limits respect the thermodynamic hierarchy: the ensemble-level bound restricts the dynamics more tightly than the average of the stochastic bounds. This statistical convergence is further evidenced by the asymptotic vanishing of the interference cross-term $\langle \tilde{f}_Q^X \rangle$ as the sample size increases. Most importantly, Fig.~\ref{fig:QSLs_driven}(d) validates that the fundamental geometric uncertainty relation $l^2 \le j$ is strictly satisfied for every individual realization, establishing the CQFI as a robust metric for single-shot thermodynamic bounds.

\begin{figure}[H]
    \centering
    \includegraphics[width=8.6cm]{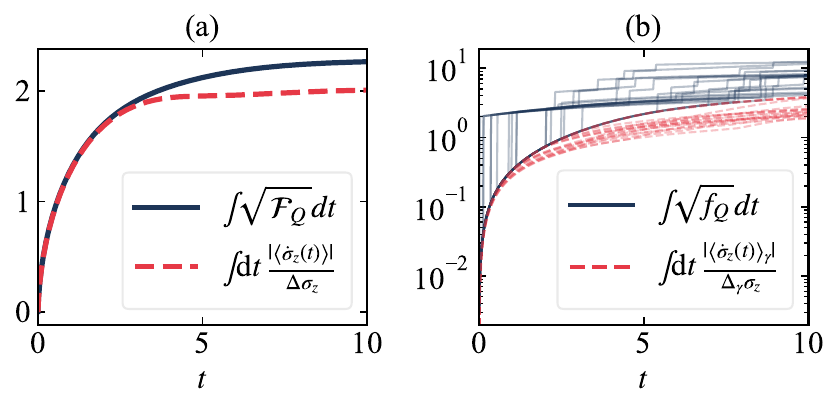}
    \caption{Speed limits for the rate of change of $\sigma_z$ expectation value. (a)  Ensemble-level speed limit [cf. Eq.\eqref{eq:Avg_QSLs_Obs_QFI}].  (b) Trajectory-level speed limit [cf. Eq.\eqref{eq:Trajs_QSLs_Obs_QFI}]. The numerical calculations evidence that the speed limits on trajectory level are respected. The inequalities are satisfied, with the geometric bounds (RHS) remaining strictly positive and bounding the zero velocity (LHS) throughout the evolution. Moreover, the single trajectory limits are numerically shown to saturate the inequality for some specific trajectories even for time regimes where the ensemble level inequality does not saturate.}
    \label{fig:QSLs_Obs}
\end{figure}

To test the speed limits of observables introduced in Eqs.~(\ref{eq:Avg_QSLs_Obs_QFI}) and (\ref{eq:Trajs_QSLs_Obs_QFI}), we take $\hat{O} = \hat{\sigma}_z$. Each bound is satisfied in the form $\int_0^t \mathrm{d}t'\,|\dot{o}(t')|/\Delta O \le 2\mathcal{L}(t)$ at the ensemble level and $\int_0^t \mathrm{d}t'\,|\dot{o}_\gamma(t')|/\Delta_{\rho_\gamma} O \le 2\ell(\gamma,t)$ per realization. We stress that the trajectory-level right-hand side is built from the trajectory's own SLD: since the conditioned state $\rho_\gamma = \ketbra{\psi_\gamma(t)}{\psi_\gamma(t)}$ is pure, $f_Q[\rho_\gamma(t)] = 4\langle\partial_t\psi_\gamma^\perp|\partial_t\psi_\gamma^\perp\rangle$ is the Fubini--Study speed squared of the trajectory, not the ensemble SLD sandwiched in $\ket{\psi_\gamma}$. The results are depicted in Fig.~\ref{fig:QSLs_Obs}: panel (a) shows the inequality at the ensemble level and panel (b) at the trajectory level for a subset of trajectories. Both bounds hold throughout the evolution, with the geometric right-hand side remaining strictly positive, and a majority of trajectories approach saturation in time regimes where the ensemble inequality does not.
\begin{figure}[H]
    \centering
    \includegraphics[width=8.6cm]{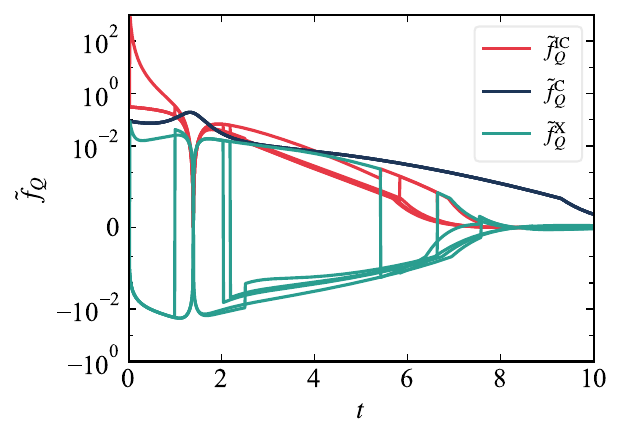}
    %\fbox{\parbox[c][3.0cm][c]{8.0cm}{\centering\textcolor{blue}{\emph{[Placeholder for new panel: insert single\_traj\_decomposition.pdf generated from cqfi.ipynb.]}}}}
    \caption{Single-trajectory decomposition of the projected SLD variance. The incoherent $\tilde{f}_Q^{\mathrm{IC}}$, coherent $\tilde{f}_Q^{\mathrm{C}}$, and cross $\tilde{f}_Q^{\mathrm{X}}$ contributions are shown along a few representative trajectories (thin lines), with their ensemble averages overlaid (thick lines). The cross-term $\tilde{f}_Q^{\mathrm{X}}$ is sign-changing along individual trajectories while $\langle\tilde{f}_Q^{\mathrm{X}}\rangle\to 0$ on averaging. At the parameters used ($\varepsilon = 0.15$, $\Gamma_0 = 0.4$, $T = 0.8$) the cross-term is comparable to $\tilde{f}_Q^X$ (within a factor $\sim 10^{-1}$ over the transient), rather than being suppressed as it is after ensemble averaging.}
    \label{fig:single_traj_decomp}
\end{figure}
Finally, Fig. \ref{fig:single_traj_decomp} shows the expected order of magnitude for the decomposition terms of the SLD projected variance, $\tilde{f}_Q^{\mathrm{IC}}$, $\tilde{f}_Q^{\mathrm{C}}$, and $\tilde{f}_Q^{\mathrm{X}}$ for $N_{{\rm traj}} = 5$. We show that for the adopted parameters, the cross term $\tilde{f}_Q^{\mathrm{X}}$ is comparable to the coherent term. 
\section{Conclusions and outlooks}
\label{secV}

We established a formal connection between stochastic thermodynamics and quantum information geometry, extending the study of metrological sensitivity from ensemble averages to individual quantum trajectories. By introducing CQFI, we have shown that at the trajectory level the information decomposes threefold: coherent, incoherent and an interference terms where the last one can assume negative values, which signals destructive interference between classical and quantum channels. The formulation allows defining the thermodynamic length and action for individual trajectories, as well as establishing single-shot velocity limits, numerically validated in a thermal qubit. In practical terms, CQFI paves the way for real-time adaptive metrology protocols, loss diagnostics in quantum engines, and sampling strategies focused on rare trajectories. Future investigations should extend the formalism to multiparametric estimations, further investigations, and experimental implementations on quantum platforms, with the goal of transforming these theoretical predictions into operational tools.

A key insight from our analysis is the spectral decomposition of the trajectory-level information. We have shown that the sensitivity of a single trajectory is driven by three distinct mechanisms: an incoherent contribution from population shifts, a coherent contribution from basis rotations, and a transient cross-term arising from their interference. Crucially, this term can be negative, a purely quantum phenomenon signifying destructive interference between classical probability updates and unitary evolution. Such negative interference acts as a strictly local witness of non-classicality, as it vanishes exactly in the ensemble average, and provides a new metric for characterizing the quality of state preparation and evolution in quantum estimation tasks.

Building on this geometric structure, we have derived fundamental quantum speed limits valid for single realizations. By defining thermodynamic length and action for individual trajectories, we have demonstrated that the speed of quantum evolution is bounded not just on average, but stochastically for every run of the experiment. These trajectory-level bounds offer a robust thermodynamic interpretation of the CQFI, connecting information-theoretic quantities to the fluctuating energetics of small quantum systems.

About the experimental accessibility, evaluating either object requires the SLD operator $L_\theta$, which is itself fixed by the state through the Lyapunov equation $\partial_\theta\rho_\theta = \tfrac12\{L_\theta,\rho_\theta\}$. In this sense the CQFI does not bypass the need to characterize the dynamics. What the projected SLD variance $\tilde{f}_{Q,\alpha}=\langle\alpha|L_\theta^2|\alpha\rangle$ avoids is reconstructing the full spectrum of $\rho_\theta$. Only the action $L_\theta|\alpha\rangle$ along the measured direction is needed. In practice $L_\theta$ may be accessed without full tomography through modulated-Hamiltonian or time-dependent perturbation-theory protocols that map the parameter derivative onto a measurable response~\cite{Hauke2016, Yu2022}, making the trajectory-resolved quantities estimable in continuously monitored platforms.

Looking forward, our framework opens several promising avenues for quantum control and thermodynamics. First, the decomposition of the CQFI offers a blueprint for real-time adaptive metrology \cite{Wiseman1995}. By monitoring the stochastic evolution of incoherent and coherent contributions, feedback protocols could dynamically switch measurement bases to exploit transient spikes in sensitivity \cite{Albarelli2018}. Concretely, a transient spike in the coherent contribution $\tilde{f}_Q^{C}$ along a trajectory flags a moment of high basis-rotation sensitivity and can serve as a real-time trigger for switching the measurement basis. Second, the ability of the cross-term to witness destructive interference suggests its usefulness in characterizing efficiency losses in microscopic heat engines or identifying quantum advantages in battery charging protocols at the single-shot level. Finally, the CQFI serves as a natural estimator for importance sampling, allowing for the efficient simulation and identification of rare, high-information trajectories that are typically inaccessible to standard Monte Carlo methods. In this role the accumulated CQFI acts as an importance-sampling weight: resampling trajectories in proportion to their integrated CQFI concentrates computational effort on the high-information realizations and reduces the variance of ensemble estimates of the QFI. These tools collectively advance our ability to analyze and optimize quantum dynamics beyond the limitations of ensemble descriptions.

\section*{Acknowledgments} PBM is grateful for discussions with Anna Sanpera, Alessandro Candeloro, Carmem Gilardoni, Diogo Soares-Pinto, Gabriel Bié Alves, Guilherme Zambon, Ivan Medina, and Luis Pedro Garc\'ia-Pintos. PBM acknowledges the support of the Brazilian agency Coordena\c{c}\~{a}o de Aperfei\c{c}oamento de Pessoal de Ensino Superior (CAPES), finance code 001. PVP acknowledges the Funda\c{c}$\Tilde{\text{a}}$o de Amparo $\Grave{\text{a}}$ Pesquisa do Estado do Rio de Janeiro (FAPERJ processes SEI-260003/005741/2024, SEI-260003/021783/2025). F.I. acknowledges financial support from the Brazilian funding agencies CAPES, CNPQ, FAPERJ (No. 151064/2022-9, and No. E-26/201.365/2022), and by the Serrapilheira Institute (grant number Serra – 2211-42166). SMDQ acknowledges CNPq grant No. 302348/2022-0, and FAPERJ (grants No. APQ1-210.310/2024, and E-26/200.155/2026). MP acknowledges support from the Royal Society Wolfson Fellowship (RSWF/R3/183013), the Department for the Economy of Northern Ireland under the US-Ireland R\&D Partnership Programme, the PNRR PE Italian National Quantum Science and Technology Institute (PE0000023), and the EU Horizon Europe EIC Pathfinder project QuCoM (GA no. 10032223). WAMM acknowledges CNPq grant No. 308560/2022-1.
\appendix

\section{Derivation of the Conditional Quantum Fisher Information (CQFI)}
\label{app:cqfi}

For an unknown parameter $\theta$, the classical Stochastic Fisher
Information (SFI) associated with a single measurement outcome $x$ is
\begin{equation}
\iota(\theta,x)=\left(\frac{\partial \log p(x|\theta)}{\partial\theta}\right)^{2}.
\label{eq:appA-sfi}
\end{equation}
In the quantum setting the conditional probability is fixed by the Born
rule, $p(x|\theta)=\Tr(\Pi_x\rho_\theta)$, where $\{\Pi_x\}$ is a POVM.
Throughout this appendix we take the elements to be orthogonal projectors,
$\Pi_x=\Pi_x^\dagger=\Pi_x^2$, so that $\sqrt{\Pi_x}=\Pi_x$; the
state-conditioned form discussed below corresponds to the rank-one choice
$\Pi_\alpha=|\alpha\rangle\langle\alpha|$.

The standard Quantum Fisher Information (QFI) is the maximum of the
classical Fisher information over all POVMs,
\begin{equation}
F_Q(\theta)=\Tr\!\big(\rho_\theta L_\theta^{2}\big)=\max_{\{\Pi_x\}} I(\theta),
\label{eq:appA-qfi}
\end{equation}
where the Symmetric Logarithmic Derivative (SLD) operator $L_\theta$ is
the Hermitian solution of the Lyapunov equation
\begin{equation}
\partial_\theta\rho_\theta=\tfrac{1}{2}\{L_\theta,\rho_\theta\}.
\label{eq:appA-lyap}
\end{equation}
Because $\rho_\theta$ and $L_\theta$ are Hermitian, cyclicity of the trace
gives $\Tr(\rho_\theta L_\theta^{2})=\Tr(L_\theta\rho_\theta L_\theta)$; we
use this identity repeatedly below.

\subsection{From the surprisal rate to the CQFI}

We seek the quantum analogue of the SFI kernel
\eqref{eq:appA-sfi}. Expanding the logarithmic derivative and inserting
the Born rule,
\begin{equation}
\iota(\theta,x)=\left(\frac{1}{p(x|\theta)}\,\frac{\partial p(x|\theta)}{\partial\theta}\right)^{2}
=\left(\frac{\Tr(\Pi_x\,\partial_\theta\rho_\theta)}{\Tr(\Pi_x\rho_\theta)}\right)^{2}.
\label{eq:appA-iota-born}
\end{equation}
Substituting the SLD identity \eqref{eq:appA-lyap}, the numerator becomes
\begin{equation}
\Tr(\Pi_x\,\partial_\theta\rho_\theta)
=\tfrac{1}{2}\Tr\!\big(\Pi_x\{L_\theta,\rho_\theta\}\big)
=\Re\,\Tr(\Pi_x L_\theta\rho_\theta),
\label{eq:appA-numer}
\end{equation}
where the last equality uses
$\Tr(\Pi_x\rho_\theta L_\theta)=\overline{\Tr(\Pi_x L_\theta\rho_\theta)}$,
valid because $\Pi_x$, $\rho_\theta$ and $L_\theta$ are Hermitian. Hence
\begin{equation}
\iota(\theta,x)=\frac{\big[\Re\,\Tr(\Pi_x L_\theta\rho_\theta)\big]^{2}}
{\big[\Tr(\Pi_x\rho_\theta)\big]^{2}}.
\label{eq:appA-iota-exact}
\end{equation}
Equation~\eqref{eq:appA-iota-exact} is exact. To obtain a quantity that
depends only on the local SLD structure---and that recovers the QFI on
averaging---we bound it in two steps. First, $[\Re z]^{2}\le|z|^{2}$ gives
\begin{equation}
\iota(\theta,x)\le\frac{\big|\Tr(\Pi_x L_\theta\rho_\theta)\big|^{2}}
{\big[\Tr(\Pi_x\rho_\theta)\big]^{2}}.
\label{eq:appA-step1}
\end{equation}
Second, we apply the Cauchy--Schwarz inequality
$|\Tr(A^\dagger B)|^{2}\le\Tr(A^\dagger A)\,\Tr(B^\dagger B)$ with the
choice
\begin{equation}
A=\sqrt{\rho_\theta}\,\sqrt{\Pi_x},\qquad
B=\sqrt{\rho_\theta}\,L_\theta\sqrt{\Pi_x}.
\label{eq:appA-AB}
\end{equation}
Using $\sqrt{\Pi_x}=\Pi_x$ and cyclicity,
\begin{align}
\Tr(A^\dagger B)&=\Tr(\Pi_x\rho_\theta L_\theta), \nonumber\\
\Tr(A^\dagger A)&=\Tr(\Pi_x\rho_\theta)=\Tr(\rho_\theta\Pi_x), \nonumber\\
\Tr(B^\dagger B)&=\Tr(\Pi_x L_\theta\rho_\theta L_\theta).
\label{eq:appA-CSterms}
\end{align}
The Cauchy--Schwarz inequality therefore yields
\begin{equation}
\big|\Tr(\Pi_x L_\theta\rho_\theta)\big|^{2}
\le \Tr(\rho_\theta\Pi_x)\,\Tr(\Pi_x L_\theta\rho_\theta L_\theta),
\label{eq:appA-CS}
\end{equation}
and combining \eqref{eq:appA-step1} with \eqref{eq:appA-CS},
\begin{equation}
\iota(\theta,x)\le
\frac{\Tr(\Pi_x L_\theta\rho_\theta L_\theta)}{\Tr(\rho_\theta\Pi_x)}.
\label{eq:appA-bound}
\end{equation}
The right-hand side saturates the local information bound and serves as
our definition of the CQFI:
\begin{equation}
\,
f_{Q,x}(\theta)=\frac{\Tr(\Pi_x L_\theta\rho_\theta L_\theta)}
{\Tr(\rho_\theta\Pi_x)}\, .
\label{eq:appA-def1}
\end{equation}
We stress that \eqref{eq:appA-def1} is the operator ordering produced by
the Cauchy--Schwarz choice \eqref{eq:appA-AB}: the central factor is
$L_\theta\rho_\theta L_\theta$, not $L_\theta^{2}\rho_\theta$. The two
coincide only when $[\rho_\theta,L_\theta]=0$ (see
Sec.~\ref{app:equality}); we keep the symmetric ordering
\eqref{eq:appA-def1} as the primary definition because it is manifestly
real and non-negative for every outcome, whereas
$\Tr(\Pi_x L_\theta^{2}\rho_\theta)/\Tr(\rho_\theta\Pi_x)$ is in general
complex for a rank-one $\Pi_x$ that is not an eigenprojector of
$\rho_\theta$.

Differently from the QFI, which is a global scalar obtained after maximizing
over outcomes, $f_{Q,x}(\theta)$ is a random variable attached to the specific
outcome $x$ of a single realization.

The non-negativity asserted above can be shown explicitly. Since $\rho_\theta\succeq 0$ admits a Hermitian square root $\sqrt{\rho_\theta}$, define $M=\sqrt{\rho_\theta}\,L_\theta$. Then the numerator of \eqref{eq:appA-def1} is
\begin{align}
\Tr(\Pi_x L_\theta\rho_\theta L_\theta)
&=\Tr\!\big(\Pi_x M^\dagger M\big)\nonumber
\\&=\Tr\!\big[(M\sqrt{\Pi_x})^\dagger (M\sqrt{\Pi_x})\big]\ge 0,
\label{eq:appA-nonneg}
\end{align}
because $L_\theta\rho_\theta L_\theta=M^\dagger M$ is positive semi-definite and $\Pi_x=\sqrt{\Pi_x}^{\,2}$. The denominator $\Tr(\rho_\theta\Pi_x)=p(x|\theta)\ge 0$ is a probability, so $f_{Q,x}(\theta)\ge 0$ for every outcome $x$.

\subsection{Recovery of the QFI on averaging}

Averaging \eqref{eq:appA-def1} over the outcome statistics
$p(x|\theta)=\Tr(\rho_\theta\Pi_x)$ removes the conditioning,
\begin{align}
\langle f_Q\rangle
&=\sum_x p(x|\theta)\,f_{Q,x}(\theta)
=\sum_x \Tr(\Pi_x L_\theta\rho_\theta L_\theta) \nonumber\\
&=\Tr\!\Big[\Big(\textstyle\sum_x\Pi_x\Big)L_\theta\rho_\theta L_\theta\Big]
=\Tr(L_\theta\rho_\theta L_\theta) \nonumber\\
&=\Tr(\rho_\theta L_\theta^{2})=F_Q(\theta),
\label{eq:appA-average}
\end{align}
where the third equality uses the completeness relation
$\sum_x\Pi_x=\mathbb{I}$ and the last uses the cyclicity identity noted
below \eqref{eq:appA-lyap}. Thus the CQFI is an unbiased,
outcome-resolved decomposition of the QFI.

\subsection{State-conditioned form}

For a rank-one projector $\Pi_\alpha=|\alpha\rangle\langle\alpha|$ it is
convenient to introduce a second, purely state-conditioned object,
\begin{equation}
\tilde f_{Q,\alpha}(\theta)=\Tr(\Pi_\alpha L_\theta^{2})
=\langle\alpha|L_\theta^{2}|\alpha\rangle,
\label{eq:appA-def2}
\end{equation}
which measures the local power of the SLD generator along a chosen
direction $|\alpha\rangle$ in Hilbert space. Constructing $L_\theta$ still requires knowledge of the state through the Lyapunov equation \eqref{eq:appA-lyap}; what estimating \eqref{eq:appA-def2} avoids is reconstructing the full spectrum of $\rho_\theta$, since only the SLD action $L_\theta|\alpha\rangle$ along the chosen direction is needed. The two definitions \eqref{eq:appA-def1} and
\eqref{eq:appA-def2} are distinct in general and agree under the condition
derived next.

\subsection{Equality condition between the two definitions}
\label{app:equality}

Write the spectral decomposition
$\rho_\theta=\sum_n p_n|n\rangle\langle n|$ and take the rank-one
$\Pi_\alpha=|\alpha\rangle\langle\alpha|$ with $c_n=\langle n|\alpha\rangle$.
Inserting resolutions of the identity in the eigenbasis, the
symmetric-ordering definition \eqref{eq:appA-def1} reads
\begin{equation}
f_{Q,\alpha}(\theta)
=\frac{\langle\alpha|L_\theta\rho_\theta L_\theta|\alpha\rangle}
{\langle\alpha|\rho_\theta|\alpha\rangle}
=\frac{\sum_{m}p_m\,\big|\langle\alpha|L_\theta|m\rangle\big|^{2}}
{\sum_n p_n\,|c_n|^{2}},
\label{eq:appA-f1-spectral}
\end{equation}
while the state-conditioned definition \eqref{eq:appA-def2} reads
\begin{equation}
\tilde f_{Q,\alpha}(\theta)
=\langle\alpha|L_\theta^{2}|\alpha\rangle
=\sum_{m}\big|\langle\alpha|L_\theta|m\rangle\big|^{2}.
\label{eq:appA-f2-spectral}
\end{equation}
Equation~\eqref{eq:appA-f1-spectral} is a $p_m$-weighted average of the
same nonnegative terms $|\langle\alpha|L_\theta|m\rangle|^{2}$ that appear
unweighted in \eqref{eq:appA-f2-spectral}, with weights normalized by
$\sum_n p_n|c_n|^{2}$. The two expressions coincide precisely when the
weighting is flat over the terms that contribute, i.e. when
\begin{equation}
p_m=\langle\alpha|\rho_\theta|\alpha\rangle
\quad\text{for every $m$ with } \langle\alpha|L_\theta|m\rangle\neq 0 .
\label{eq:appA-condition}
\end{equation}
This is the general equality condition. It is met in the two physically
relevant situations below.

\paragraph*{(i) Measurement in an eigenbasis of $\rho_\theta$.}
If $|\alpha\rangle=|k\rangle$ is an eigenstate of $\rho_\theta$ then
$\langle\alpha|\rho_\theta|\alpha\rangle=p_k$ and
\eqref{eq:appA-f1-spectral} becomes
$\sum_m (p_m/p_k)\,|\langle k|L_\theta|m\rangle|^{2}$. This equals
\eqref{eq:appA-f2-spectral} iff $p_m=p_k$ for all $m$ coupled to $k$ by
$L_\theta$, i.e. $L_\theta|k\rangle$ has support only inside the
eigenspace of $p_k$.

\paragraph*{(ii) Commuting case $[\rho_\theta,L_\theta]=0$.}
When the SLD commutes with the state, $L_\theta\rho_\theta L_\theta
=L_\theta^{2}\rho_\theta=\rho_\theta L_\theta^{2}$ as operators, and
$L_\theta$ is block-diagonal in the eigenbasis of $\rho_\theta$, so the
only nonvanishing terms in \eqref{eq:appA-f1-spectral} have $p_m=p_k$.
Condition~\eqref{eq:appA-condition} is then satisfied for every
eigenprojector and
\begin{equation}
f_{Q,\alpha}(\theta)=\tilde f_{Q,\alpha}(\theta).
\end{equation}
This is the incoherent (classical) regime: the coherent and cross terms of
the spectral decomposition (Appendix~B) vanish, the CQFI reduces to the
classical SFI of \eqref{eq:appA-sfi}, and the two definitions agree. It
corresponds to quasi-static or adiabatic evolutions in which the system
remains in an instantaneous eigenstate. The Gaussian force-sensing example
of Appendix~C\,2 furnishes a further instance in which both definitions
coincide and, moreover, become independent of the outcome $\alpha$,
because the SLD is linear in the quadratures and the measurement shifts
only the first moments of the state.

Outside these regimes the symmetric ordering \eqref{eq:appA-def1} is the
correct outcome-resolved object: it is real and non-negative, saturates the
local bound \eqref{eq:appA-bound}, and averages to the QFI via
\eqref{eq:appA-average}.

\section{Spectral Decomposition of $\tilde{f}_{Q,\alpha}(\theta)$ \label{App:B}}

To elucidate the physical mechanisms contributing to the CQFI, we assume the density matrix possesses the spectral decomposition $\rho_\theta = \sum_n p_n(\theta) \ketbra{n_\theta}{n_\theta}$. In this eigenbasis, the SLD operator can be expanded as
\begin{equation}
    L_{\theta} = \sum_{n}\frac{\partial_{\theta}p_{n}}{p_{n}}\ketbra{n_{\theta}}{n_{\theta}} + 2\sum_{n \neq k}\frac{p_n - p_{k}}{p_{n} + p_{k}}\braket{n_{\theta}}{\partial_{\theta}k_{\theta}}\ketbra{n_{\theta}}{k_{\theta}}.
\end{equation}
We can decompose the SLD into a diagonal (incoherent) component $L_{IC}$ and an off-diagonal (coherent) component $L_{C}$, such that $L_\theta = L_{IC} + L_{C}$. The squared operator is then given by $L_\theta^2 = L_{IC}^2 + L_{C}^2 + \{L_{IC}, L_{C}\}$.

When evaluating $\tilde{f}_{Q,\alpha}(\theta) = \langle \alpha | L_\theta^2 | \alpha \rangle$, we must account for the cross-terms between the diagonal and off-diagonal parts. While these cross-terms vanish in the global ensemble average (the trace), they are generally non-zero for a specific state $|\alpha\rangle$. Thus, the projected SLD variance splits into three distinct contributions
\begin{equation}
    \tilde{f}_{Q,\alpha}(\theta) = \tilde{f}_{Q,\alpha}^{IC}(\theta) + \tilde{f}_{Q,\alpha}^{C}(\theta) + \tilde{f}_{Q,\alpha}^{X}(\theta).
\end{equation}

The first term, the incoherent CQFI, is analogous to the classical SFI. It captures the sensitivity arising from changes in the state's eigenvalues
\begin{equation}
    \tilde{f}_{Q,\alpha}^{IC}(\theta) = \sum_n \abs{\braket{n}{\alpha}}^2 \left(\frac{\partial_\theta p_n}{p_n}\right)^2.
\end{equation}

The second term, the coherent CQFI $\tilde{f}_Q^{C}$, captures the information arising from the unitary rotation of the eigenbasis. This contribution reflects the fact that the eigenstates depend on $\theta$ and generally do not commute with their derivatives
\begin{equation}
    \tilde{f}_{Q,\alpha}^{C}(\theta) = \sum_k \left|\sum_{n (\neq k)} \braket{n}{\alpha} \frac{2(p_n - p_k)}{p_n + p_k} \braket{k}{\partial_\theta n}\right|^2.
\end{equation}

The third term, the cross-term CQFI $\tilde{f}_Q^{X}$, represents the interference between the population dynamics and the basis rotations
\begin{align}
    \tilde{f}_{Q,\alpha}^{X}(\theta) &= \sum_{k \neq n} \text{Re}\left[ c_{k}^{*}c_n \left(\frac{\partial_\theta p_k}{p_k} + \frac{\partial_\theta p_n}{p_n}\right)\right.\\&\times\left. \left(\frac{2(p_n - p_k)}{p_n + p_k}\right) \braket{k}{\partial_\theta n} \right],
\end{align}
where $c_n = \braket{n}{\alpha}$. This term quantifies the correlation between the classical and quantum channels of information. Notably, unlike the incoherent and coherent contributions which are strictly non-negative, the cross-term $\tilde{f}_Q^X$ can be negative. A negative value indicates destructive interference, where the population shifts and geometric rotations partially cancel each other out relative to the probe state $|\alpha\rangle$.

The cross-term vanishes on averaging over the measurement statistics. With $\Pi_\alpha=\ketbra{\alpha}$ and $c_n=\braket{n}{\alpha}$, averaging $c_k^* c_n=\braket{\alpha}{k}\braket{n}{\alpha}$ over a complete set of outcomes uses $\sum_\alpha \ketbra{\alpha}=\mathbb{I}$, so that $\sum_\alpha c_k^* c_n=\braket{n}{k}=\delta_{nk}$. Since the cross-term \eqref{eq:CQFI_X} runs strictly over $k\ne n$, every contribution is multiplied by $\delta_{nk}=0$, and hence
\begin{equation}
\langle \tilde{f}_Q^X\rangle=\sum_\alpha p(\alpha|\theta)\,\tilde{f}_{Q,\alpha}^X(\theta)\Big|_{\text{completeness}}=0,
\end{equation}
recovering the standard two-term ensemble decomposition of the QFI.

\section{Examples for CQFI}

In order to put in place the formal framework illustrated above and attest to the potential of CQFI, we examine two paradigmatic examples: field-sensing with a qubit prepared in a thermal state, and force-sensing through displaced Gaussian states. We demonstrate that the CQFI behaves qualitatively similarly to the QFI, and in specific Gaussian regimes, they coincide quantitatively.

\subsection{Field sensing with a thermal qubit}

We consider a two-level system used to estimate a transverse field parameter $\theta$. The system Hamiltonian is $\hat{H}_S(\theta) = \Delta\sigma_z + \frac{\theta}{2}\sigma_x$, where $\Delta$ is the energy gap and $\theta$ is the transverse driving field. The system is in a thermal state $\rho_\theta = e^{-\beta \hat{H}_S(\theta)}/\mathcal{Z}(\theta)$, where $\mathcal{Z}(\theta) = \mathrm{Tr}(e^{-\beta \hat{H}_S(\theta)})$.

The eigenvalues of the Hamiltonian are $E_{\pm}(\theta) = \pm \frac{\Omega(\theta)}{2}$, with the generalized Rabi frequency $\Omega(\theta) = \sqrt{\Delta^2 + \theta^2}$. The equilibrium populations are
\begin{equation}
    p_\pm(\theta) = \frac{1}{2} \left[1 \mp \tanh\left(\frac{\beta\Omega(\theta)}{2}\right)\right].
\end{equation}
The incoherent contribution to the CQFI, arising from the population derivatives, is
\begin{equation}
    f_{Q,\Pi_\pm}^{\mathrm{IC}}(\Pi_\pm, \theta) = \frac{\beta^2 \theta^2}{4\Omega^2(\theta)} \left[1 \pm \tanh\left(\frac{\beta\Omega(\theta)}{2}\right)\right]^2.
\end{equation}
The coherent contribution, arising from the misalignment between the field and the quantization axis, is given by
\begin{equation}
    f_{Q,\Pi_\pm}^{\mathrm{C}}(\theta) = \frac{\Delta^2}{\Omega^4(\theta)}\tanh\left(\frac{\beta\Omega(\theta)}{2}\right).
\end{equation}
Note that $f_Q^{\mathrm{C}}$ vanishes only if $\beta = 0$ (infinite temperature) or $\Delta = 0$ (no gap). The total CQFI is the sum of these two terms. This example illustrates how the CQFI allows us to distinguish between information gained from thermal population shifts versus information gained from the rotation of the energy eigenbasis.

\subsection{Force sensing in dislocated Gaussian states}

\begin{figure}[!ht]
    \centering
    \includegraphics[width=1.0\linewidth]{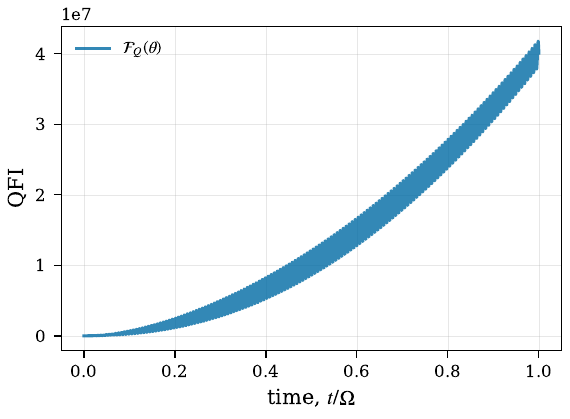}
    \caption{Time evolution of the CQFI and QFI for force sensing in dislocated Gaussian states. Due to the Gaussian nature of the state and measurement, the CQFI coincides with the QFI. Both quantities scale quadratically with time, $\mathcal{F}_{Q} \sim t^2$.}
    \label{fig:Ex2_CQFI}
\end{figure}

Dislocated Gaussian states are a prominent resource in quantum metrology, particularly for weak force estimation. We consider a harmonic oscillator under a constant force $\theta$, described by the Hamiltonian $\hat H = \hbar \omega \hat a^\dagger \hat a - \theta \hat x$.

Gaussian systems are fully characterized by their first and second moments. The optimal SLD operator for estimating the displacement parameter $\theta$ is given by~\cite{monras2013phasespaceformalismquantum, chang2025multiparameterquantumestimationgaussian}
\begin{equation}
    \hat L_\theta = \frac{2t}{\det(V)}\left[ V_{xp}(\hat x - \langle \hat x \rangle) - V_x(\hat p - \langle \hat p \rangle) \right],
\end{equation}
where $V$ is the covariance matrix. The resulting QFI is
\begin{equation}
    \mathcal{F}_Q(\theta) = \mathrm{Tr}(\rho \hat L_\theta^2) = \frac{4t^2 V_x}{\det(V)}.
\end{equation}

To compare this bound with a trajectory-level quantity, we consider a continuous weak measurement of the position quadrature, described by Gaussian POVM elements
\begin{equation}
    \hat \Pi_x(\alpha) = \left( \frac{4 k \Delta t}{\pi}\right)^{1/4} \exp\left(-2k \Delta t (\hat x-\alpha)^2\right), \label{eq:povm_gaussian}
\end{equation}
where $\alpha$ is the measurement outcome. The CQFI for this measurement is defined as
\begin{equation}
    f_{Q,\alpha}(\theta) = \frac{\mathrm{Tr}\left(\rho \hat \Pi_x(\alpha) \hat L_\theta^2 \right)}{\mathrm{Tr}\left(\rho \hat \Pi_x(\alpha)\right)}.
\end{equation}
Remarkably, solving the Gaussian integrals reveals that
\begin{equation}
    f_{Q,\alpha}(\theta) = \frac{4t^2 V_x}{\det(V)} = \mathcal{F}_Q(\theta). 
\end{equation}
The CQFI is independent of the stochastic measurement outcome $\alpha$. This is a consequence of the Gaussian statistics: while the measurement outcome $\alpha$ shifts the center of the Wigner function (updating the first moments), the curvature (second moments) and the sensitivity to the displacement force remain invariant. Thus, in this specific Gaussian regime, the single-trajectory information is identical to the ensemble average.

\providecommand{\newblock}{}
\bibliography{refs}

\end{document}